\documentclass[final,1p,times]{elsarticle}
\usepackage{graphics}
\usepackage{epsfig}

\usepackage{amssymb}
\usepackage{amsthm}
\usepackage{subfig}
\usepackage{fixltx2e}

\biboptions{numbers,sort&compress}

\journal{Physics Letters B}

\begin{document}

\begin{frontmatter}



\title{A simulation toolkit for electroluminescence assessment in rare event experiments}

\author[aveiro]{C. A. B. ~Oliveira}
\ead{carlos.oliveira@ua.pt}
\author[cern]{H. ~Schindler}
\author[cern]{R. ~Veenhof}
\author[liverpool]{S. ~Biagi}
\author[coimbra]{C. M. B. ~Monteiro}
\author[coimbra]{J. M. F. ~dos Santos}
\author[aveiro]{A. L. ~Ferreira}
\author[aveiro]{J. F. C. A. ~Veloso}

%\author{}

%\address{}

\address[aveiro]{I3N - Physics Department, University of Aveiro, 3810-193 Aveiro, Portugal}
\address[cern]{CERN, Geneva, Switzerland}
\address[liverpool]{Physics Department, University of Liverpool, Liverpool, UK}
\address[coimbra]{GIAN, Physics Department, University of Coimbra, 3004-516 Coimbra, Portugal}

\begin{abstract}
A good understanding of electroluminescence is a prerequisite when optimising double-phase noble gas detectors for Dark Matter searches and high-pressure xenon TPCs for neutrinoless double beta decay detection.

A simulation toolkit for calculating the emission of light through electron impact on neon, argon, krypton and xenon has been developed using the Magboltz and Garfield programs. Calculated excitation and electroluminescence efficiencies, electroluminescence yield and associated statistical fluctuations are presented as a function of electric field. Good agreement with experiment and with Monte Carlo simulations has been obtained. 

\end{abstract}

\end{frontmatter}

% \linenumbers

\section{Introduction}
\label{sec:intro}
Several experiments in astrophysics and cosmology, such as direct Dark Matter searches \cite{darkmatter1,darkmatter2,darkmatter3} and neutrinoless double beta decay \cite{next,exo} , are based on noble gas and/or liquid Time Projection Chambers (TPC) which use electroluminescence for primary ionisation signal amplification. This technique gives high gains and good energy resolution, and is suitable for experiments with low event rates and high background levels.

Up to now, for calculating the light yield and efficiencies, a three-dimensional Monte Carlo program of the electron drift in xenon and xenon-neon mixtures \cite{3dxe,xe-ne} and a one-dimensional program for krypton and argon \cite{1d} existed. Although validated, they are not open-source nor freely accessible, unlike the toolkit described in this paper. Our toolkit calculates the yield and efficiencies and provides a versatile and comprehensive parameterisation of the physics processes involved in electroluminescence. It uses an up-to-date database of electron-atom and electron-molecule cross sections for about 60 gases.

In this paper, calculations for neon, argon, krypton and xenon in uniform fields are discussed. We compare our results with earlier Monte Carlo simulations \cite{3dxe,1d,xe-ne} and with experimental data \cite{expxe,expar,GSPC-thvdias}.  

\section{Electroluminescence}
\label{EL}
In a gaseous detector, primary electrons are produced along the track of a incoming particle interacting in the absorption region. They are driven by an electric field, below the scintillation threshold, to a region where, under the influence of stronger fields, they can excite or ionise the gas atoms. 

\begin{figure*}
\centering
\includegraphics[width=.9\textwidth]{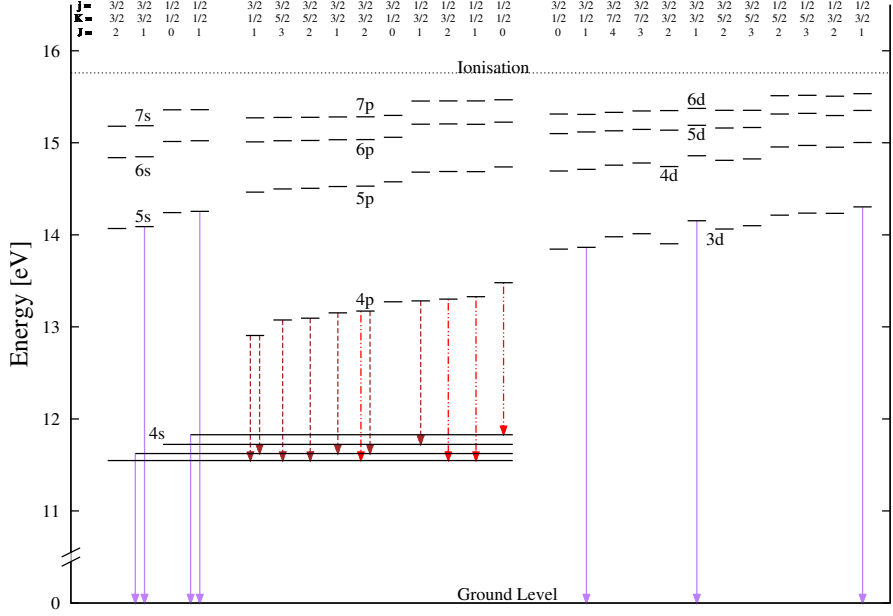}
\caption{Simplified energy level diagram of argon. The energy levels were taken from \cite{levels_ar} and the ionisation thresholds from \cite{ion}. The stronger dipole-allowed transitions are also shown \cite{persistant2}. Continuous arrows (purple) correspond to VUV, dash-dotted arrows (red) to visible and dashed arrows (brown) to IR photons.}
\label{fig:levels_ar}
\end{figure*}

The excited atomic levels are described by the $jl$ coupling \cite{jl1,jl2}. In the Racah notation, e.g. the lowest four excited levels are $n\mathrm{s} \left[ 3/2 \right]_2^\mathrm{o}$,  $n\mathrm{s} \left[ 3/2 \right]_1^\mathrm{o}$, $n\mathrm{s'} \left[ 1/2 \right]_0^\mathrm{o}$  and  $n\mathrm{s'} \left[ 1/2 \right]_1^\mathrm{o}$. Fig .\ref{fig:levels_ar} shows the level diagram of argon with the stronger dipole-allowed transitions \cite{persistant1}.

In addition to the transitions shown in Fig. \ref{fig:levels_ar}, experimental spectra of pure noble gases show a continuum in discharges and in proportional scintillation \cite{suzuki,tanaka1,tanaka2} due to excimer decay. Excimers -- electronically excited molecules, $R_2^{**}$ -- are formed through three-body collisions between an excited atom, $R^*$, and two atoms in the ground state, $R$:

\begin{equation}
\label{eq:excimerform}
R^*+2R\rightarrow R_2^{**}+R\textrm{.}
\end{equation}

Three-body collisions dominate at pressures above a few tens of Torr making the excimer formation the main channel of de-population of excited atoms \cite{p2}. In this paper we assume that other processes, e. g. associative ionisation which affects highly excited states that are not frequently produced, contribute negligibly \cite{associative1,associative2}.  

The excimers involved are mainly $^1\Sigma_\mathrm{u}^+$ and $^3\Sigma_\mathrm{u}^+$ \cite{mulliken,koehler}. They are formed through process (\ref{eq:excimerform}) in high vibrational states and can decay to the repulsive ground state, $^1\Sigma_\mathrm{g}^+$, emitting a VUV photon,

\begin{equation}
\label{eq:1stcont}
R_2^{**}\rightarrow2R+h\nu_1\textrm{,}
\end{equation}
or they can collide with ground state atoms losing vibrational energy:

\begin{equation}
\label{eq:vib_relax}
R_2^{**}+R\rightarrow R_2^*+R\textrm{.}
\end{equation}

In the latter case, the resultant excimer in a low vibrational state, $ R_2^*$ , emits a VUV photon with slightly lower energy:

\begin{equation}
\label{eq:2ndcont}
R_2^{*}\rightarrow2R+h\nu_2\textrm{.}
\end{equation}

The electronic transitions of excimers follow the Franck-Condon principle \cite{fc1,fc2}. Taking into account that the ground state is repulsive, the principle explains the continuum spectra observed experimentally for low pressures ($<100\textrm{ Torr}$) which exhibit two peaks, the ``first continuum'' at higher frequencies - and the ``second continuum'' at lower frequencies. 

At high pressures, typically above $300\textrm{ Torr}$, the proportional electroluminescence spectra show only the second continuum because process (\ref{eq:vib_relax}) dominates over process (\ref{eq:1stcont}) \cite{suzuki}.

\section{Simulation toolkit}
\label{sec:toolkit}
The simulation toolkit was developed using the new C++ version of the microscopic technique of Garfield \cite{garf++}, which currently uses Magboltz 8.9.3 \cite{mag}. 

\subsection{Garfield}
\label{subsec:garf}

Garfield is a program for the detailed simulation of gaseous detectors \cite{garf_ex}. Its Monte Carlo microscopic technique tracks electrons in gases at molecular level using procedures and cross-sections available in Magboltz.

In noble gases each collision is classified as elastic, excitation or ionisation. 

Information about each excited atom is available: the $\left(x,y,z\right)$ position, the time of production and the excitation level. This information is used to determine the electroluminescence signal properties.

The program simulates virtually any electric and magnetic field \cite{nebem}.
\begin{figure*}
\centering
  \centering
  \subfloat{\includegraphics[height=0.5\textwidth]{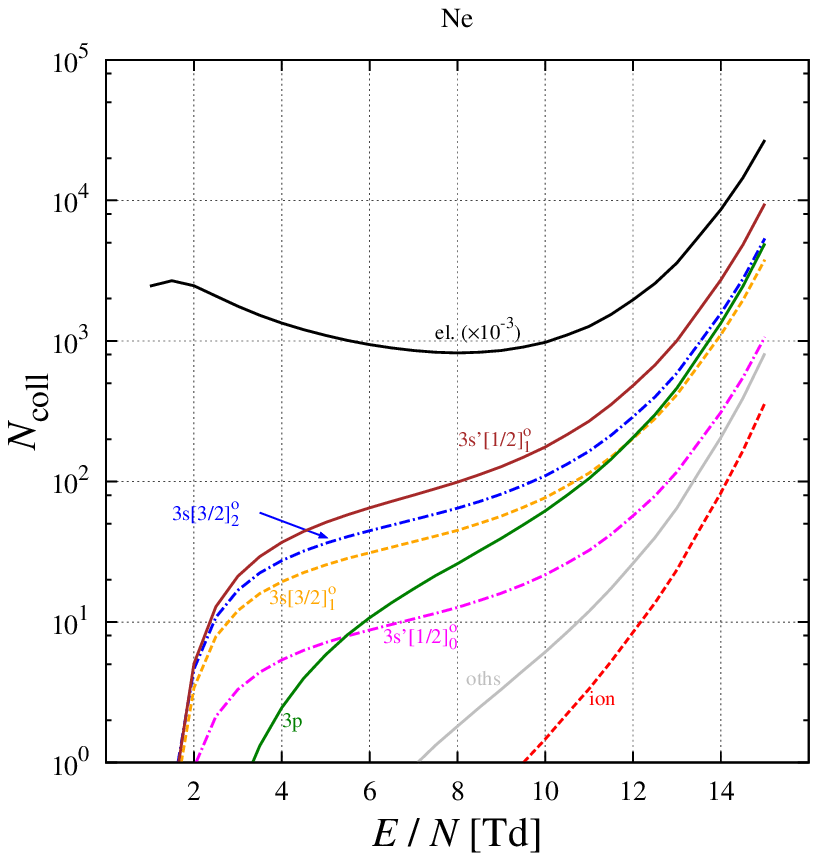}}\hspace{0.5cm}
  \subfloat{\includegraphics[height=0.5\textwidth]{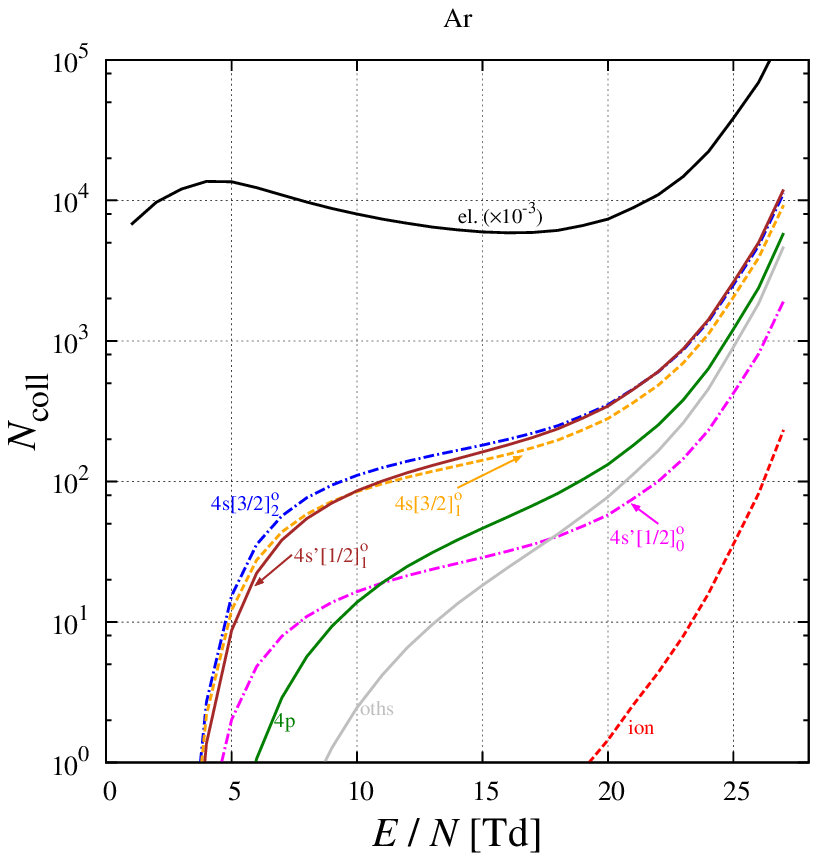}}\\
  \subfloat{\includegraphics[height=0.5\textwidth]{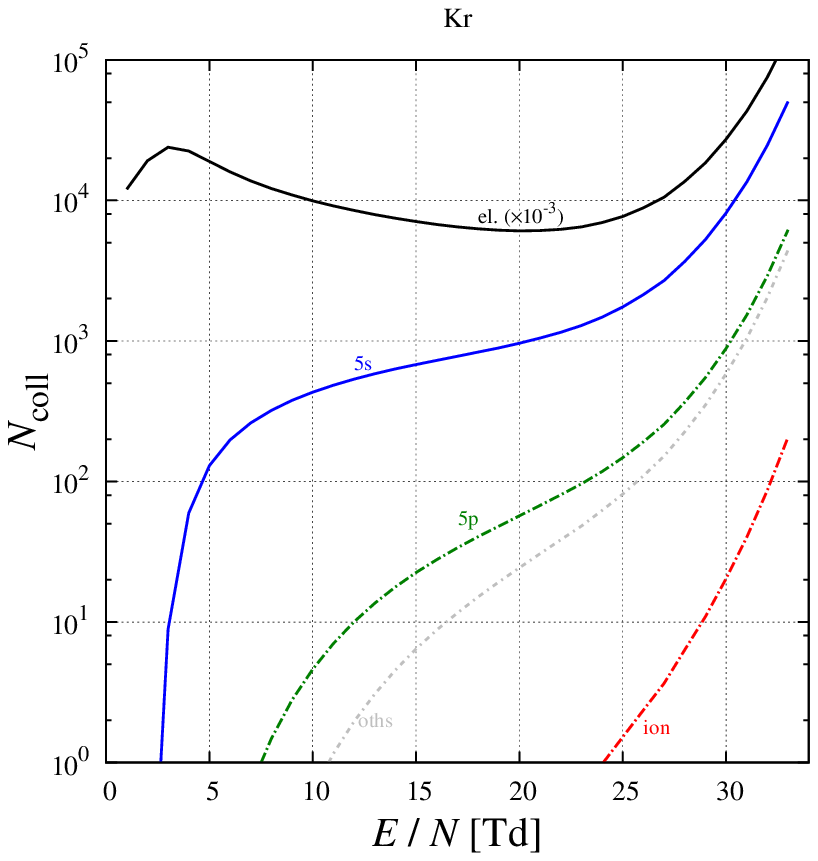}}\hspace{0.5cm}
  \subfloat{\includegraphics[height=0.5\textwidth]{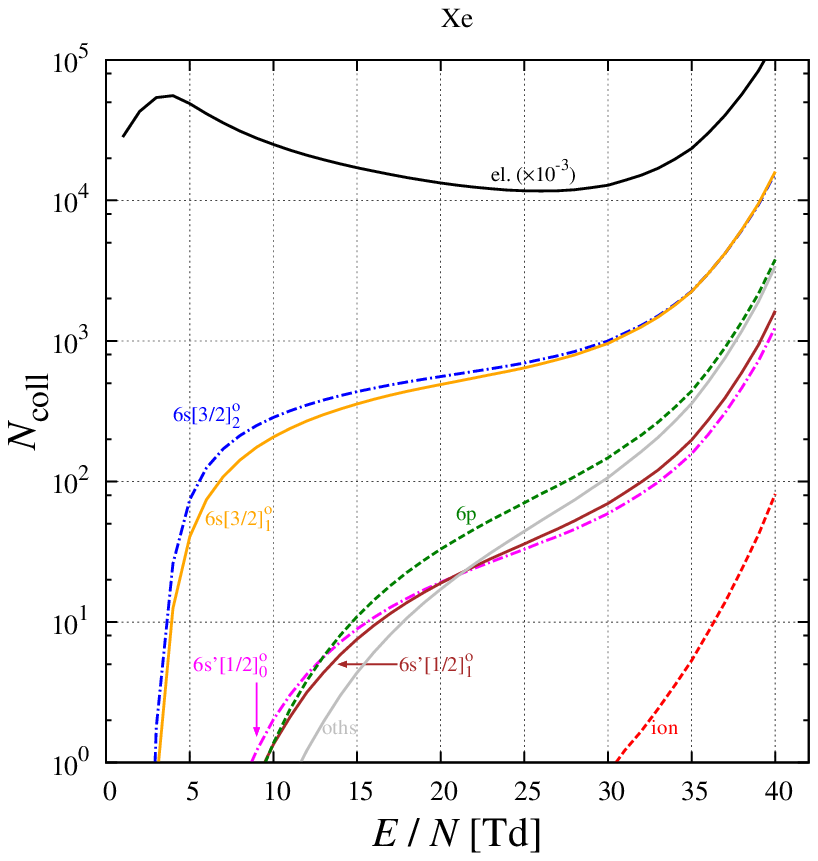}}
\caption{The total number of collisions (elastic, ionisations and of some excitation groups/levels) during electron drift over $z=2\textrm{ cm}$ of gas at $293\textrm{ K}$ and $1\textrm{ atm}$, including those experienced by secondary charges when ionisation is possible.}
\label{fig:xsid}
\end{figure*}

\subsection{Magboltz}
\label{subsec:mag}

Magboltz was developed to calculate the transport parameters of electrons drifting in gases under the influence of electric and magnetic fields \cite{mag_paper}.

For this purpose, the program contains, for 60 gases, electron cross sections for all relevant interactions. Separate excitation cross sections are available for 45, 44, 4 and 50 levels/groups of neon, argon, krypton and xenon, respectively. 

Fig. \ref{fig:xsid}  presents the total number of collisions  of different types produced during electron drift over $z=2\textrm{ cm}$ of gas at $293\textrm{ K}$ and $1\textrm{ atm}$ -- the conditions of our simulations -- as a function of the reduced electric field, $E/N$, i.e. the electric field $E$ divided by the gas number density $N$. The $n$s states represent $\sim$90 \% of all excitations in the proportional electroluminescence region (see Fig. \ref{fig:yon}) in accordance with \cite{suzuki}. The $n$p states represent between $5-10\textrm{ \%}$, depending on the gas and on the intensity of the field. 

The most frequently produced excited state in Ar and Xe is the metastable level $n\mathrm{s\left[3/2\right]_2^o}$ with a lifetime of seconds \cite{Xe-meta}, closely followed by the radiative $n\mathrm{s}$ states, with an intrinsic lifetime of nanoseconds \cite{Ar-rad,Xe-rad}. Atomic transitions from the latters are promptly re-absorbed by ground state  atoms through the radiation trapping mechanism. The long effective lifetime of these states enhances the importance of excimers.

\begin{figure}
\centering
\includegraphics[width=0.7\textwidth]{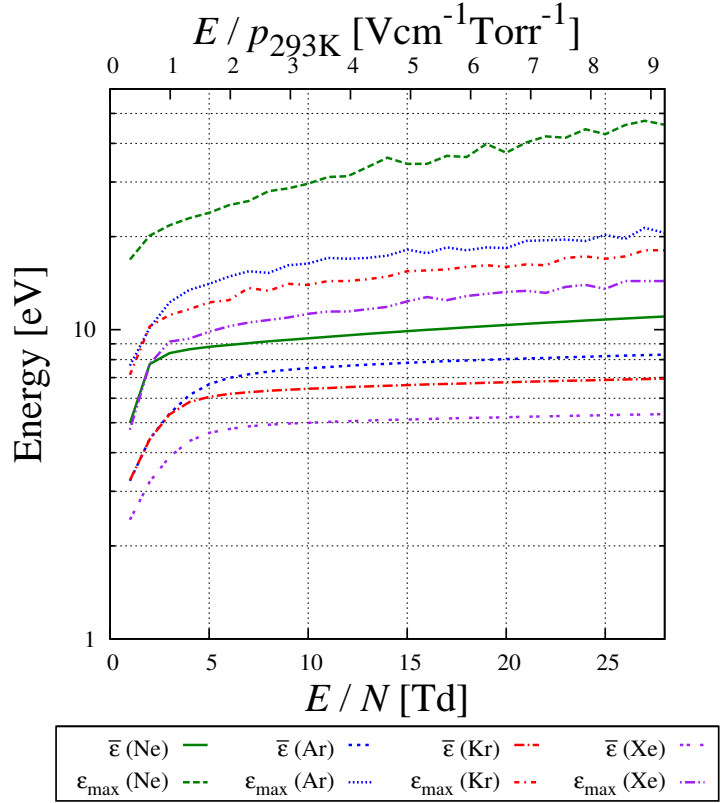}
\caption{Mean and maximum energies, $\bar{\varepsilon}$ and $\varepsilon_\mathrm{max}$ respectively, reached by the electrons before collisions, as functions of $E/N$.}
\label{fig:en_dist_analisys}
\end{figure}

Fig. \ref{fig:en_dist_analisys} shows the maximum energy of electrons, over $10^9$  collisions, as well as their mean energy. 

\section{Model}
\label{model}

We assume that each excited atom gives rise to the emission of one VUV photon from second continuum excimer decay. Its wavelength was generated according to a Gaussian distribution with mean $82.2\textrm{ nm}$, $128\textrm{ nm}$, $147\textrm{ nm}$, and $173\textrm{ nm}$ and a FWHM of $3\textrm{ nm}$, $10\textrm{ nm}$, $12\textrm{ nm}$ and $14\textrm{ nm}$ for neon, argon, krypton and xenon, respectively \cite{suzuki,tanaka1,tanaka2,morozov}. In addition, when charge multiplication occurs, each ion also gives rise to the emission of one VUV photon \cite{saito}.

The number of primary electrons for each $E/N$ was varied between $3\times 10^4$ and $2\times 10^5$ to keep the error in $J$ (see Eq. (\ref{eq:j})) below $2\textrm{ \%}$. The starting direction of each primary electron was sampled isotropically. The starting energy was distributed according to the energy distribution calculated by Magboltz.

\section{Results and discussion}
\label{results}

\begin{figure}
\centering
\includegraphics[width=0.7\textwidth]{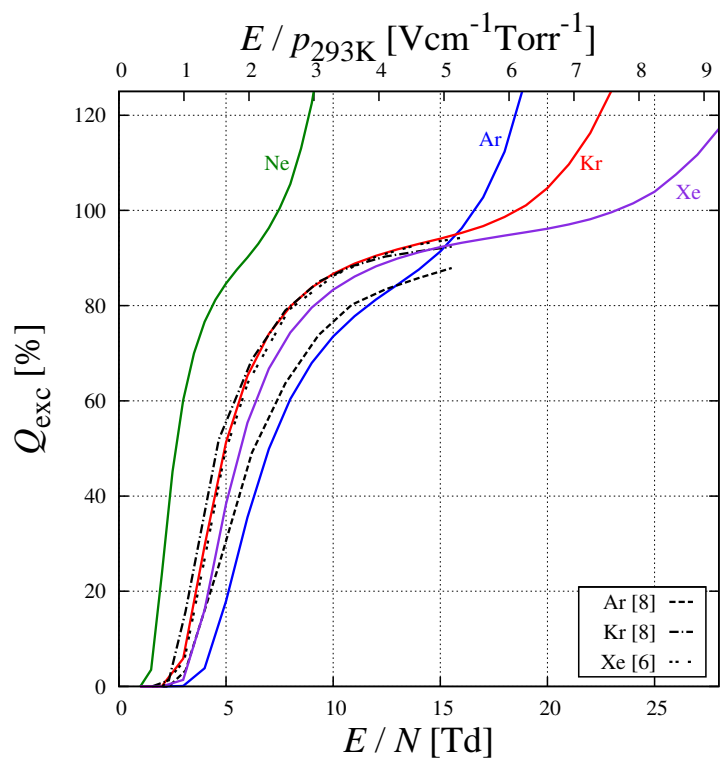}
\caption{Excitation efficiency, $Q_\mathrm{exc}$, as a function of $E/N$, compared with earlier work \cite{3dxe,1d}.}
\label{fig:qexc}
\end{figure}

Fig. \ref{fig:qexc} shows the excitation efficiency, $Q_\mathrm{exc}$, the fraction of energy acquired by the $N_\mathrm{e}$ primary electrons in the electric field that is spent in the excitation process \cite{3dxe}, 

\begin{equation}
\label{eq:qexc}
 Q_\mathrm{exc}=\frac{\displaystyle\sum_{i=1}^{i=n_\mathrm{exc}}n^i\varepsilon_\mathrm{exc}^i}{ezN_\mathrm{e}E}
\end{equation}
where $n_\mathrm{exc}$ is the number of excitation groups available in Magboltz for the studied gas, $n^i$ the number of excitations of the $i^\mathrm{th}$ group produced by the primary electrons, $\varepsilon_\mathrm{exc}^i$ the energy of the excitation group, $z$ the distance travelled by the electrons, $E$ the electric field and $e$ the elementary charge.

\begin{figure}[t]
\centering
\includegraphics[width=0.7\textwidth]{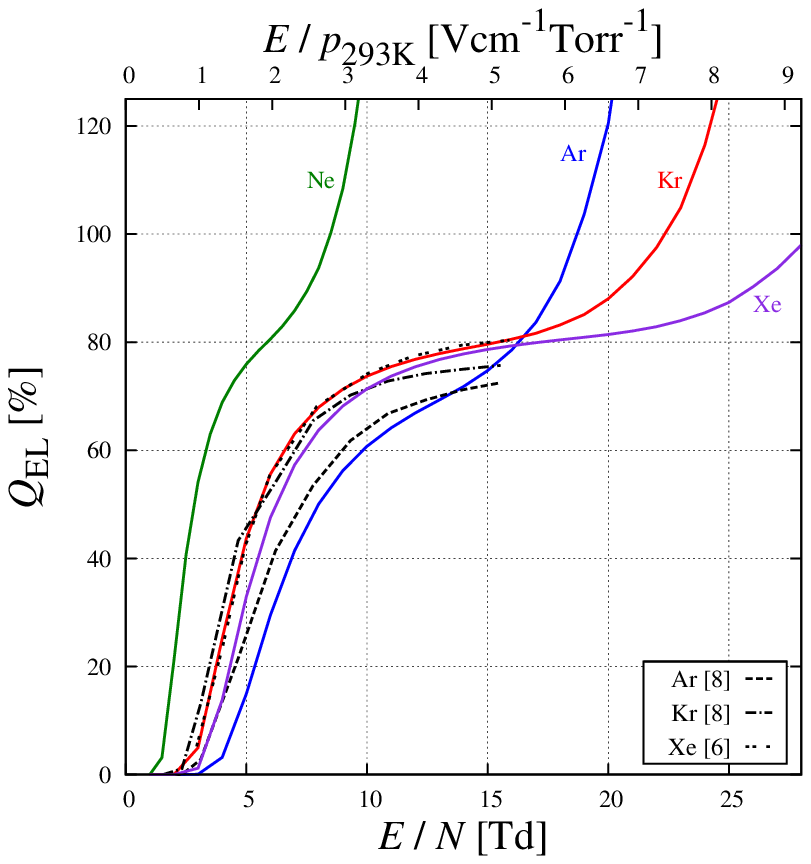}
\caption{Electroluminescence efficiency, $Q_\mathrm{EL}$, as a function of $E/N$, compared with earlier work \cite{3dxe,1d}.}
\label{fig:qel}
\end{figure}

Fig. \ref{fig:qel} shows the electroluminescence efficiency, $Q_\mathrm{EL}$, the ratio between the energy emitted in the form of VUV photons and the energy acquired by the electrons during drift:

\begin{equation}
\label{eq:qel}
 Q_\mathrm{EL}=\frac{\displaystyle\sum_{i=1}^{i=n_\mathrm{exc}}\sum_{j=1}^{j=n^i}\varepsilon_\mathrm{EL}^{i,j}}{ezN_\mathrm{e}E}
\end{equation} 
where $\varepsilon_\mathrm{EL}^{i,j}$ is the energy of the de-excitation photon of the $j^\mathrm{th}$ excited atom in the $i^\mathrm{th}$ group through excimer decay. This pa\-ra\-me\-ter was generated randomly according to the characteristics of the second continuum (Section \ref{model}).

When the electric field supplies enough energy to electrons for excitation, the cross section increases abruptly and electroluminescence thus begins with a fast increase of $Q_\mathrm{exc}$ and $Q_\mathrm{EL}$. Increasing $E/N$, a plateau is reached before $Q_\mathrm{exc}$ and $Q_\mathrm{EL}$ increase due to the electroluminescence produced by the additional secondary electrons.

$Q_\mathrm{EL}$ is always lower than $Q_\mathrm{exc}$. This is due to the loss of vibration energy from excimers before they emit a VUV photon, to radiative transitions from molecular levels to $^1\Sigma_\mathrm{u}^+$ and $^3\Sigma_\mathrm{u}^+$ and to infrared losses. 

$Q_\mathrm{exc}$ and $Q_\mathrm{EL}$ in argon and krypton agree for high $E/N$ with \cite{3dxe,1d}. Below $6$ to $8\textrm{ Td}$ ($1\textrm{ Td}=10^{-17}\textrm{ Vcm}^2$) our values are $10\textrm{ \%}$ lower. The difference increases with decreasing $E/N$. This region is not interesting for operating detectors based on electroluminescence amplification because of the large statistical fluctuations and the low light gains. 

\begin{figure}[t]
\centering
\includegraphics[width=0.7\textwidth]{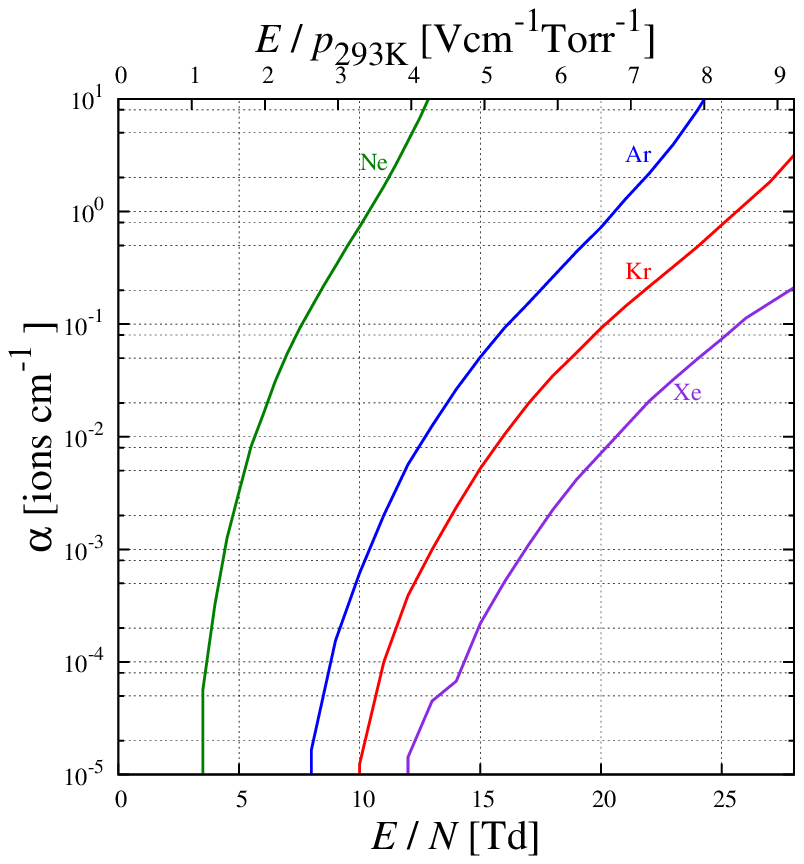}
\caption{Calculated First Townsend coefficient, $\alpha$, as a function of $E/N$ at $293\textrm{ K}$ and $1\textrm{ atm}$.}
\label{fig:alpha}
\end{figure}

The first Townsend coefficient $\alpha$, obtained dividing the number of ionisations per primary electron by the drift distance, is shown in Fig. \ref{fig:alpha}. 

\begin{figure}[t]
\centering
\includegraphics[width=0.8\textwidth]{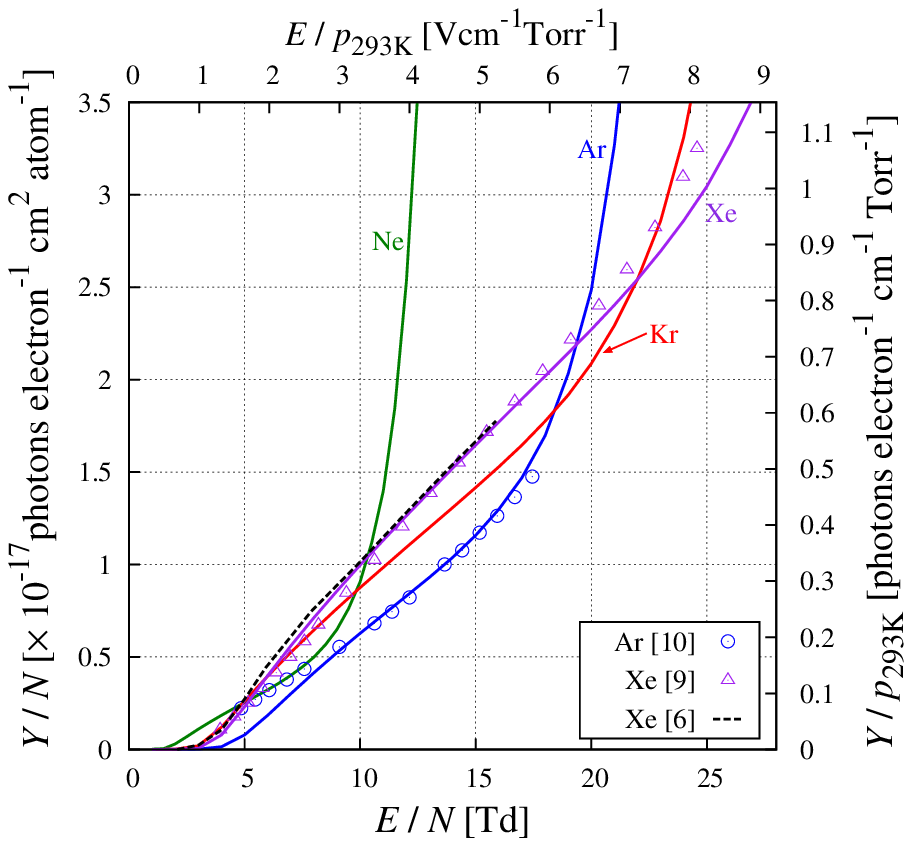}
\caption{Reduced electroluminescence yield, as a function of $E/N$, compared with earlier Monte Carlo simulation data for xenon \cite{3dxe} and measurements for argon and xenon \cite{expar,expxe}}
\label{fig:yon}
\end{figure}

Fig. \ref{fig:yon} shows the reduced electroluminescence yield $Y/N$, defined as the number of photons emitted per primary electron and per unit of drift path divided by $N$. Fig. \ref{fig:yon} also presents results of earlier simulation work for xenon \cite{3dxe} and measurements for argon and xenon \cite{expar,expxe}. The overall agreement consolidates the assumption that the main channel of de-population of excited atoms is through excimer decay. If other processes contribute they also lead to the emission of a VUV photon. The exception in the agreement is argon below $8\textrm{ Td}$ and is under investigation.

$Y/N$ is linear in $E/N$ at low $E/N$, even when some ionisation is produced (see Fig. \ref{fig:alpha}). Above $\sim$7 Td, $\sim$15 Td, $\sim$18 Td and $\sim$22 Td for neon, argon, krypton and xenon, secondary charges producing electroluminescence, change the linear behaviour of $Y/N$. These thresholds are in good agreement with those calculated for neon and xenon in Fig.5 of \cite{GSPC-thvdias}. For lower values of $E/N$, the probability of ionisation is too low for changes in the linear behaviour of $Y/N$ to be detected. The slopes increase from the lighter to the heavier gas and reflect the decrease in the minimum energy required to produce one excitation. Xenon is the gas that gives the highest electroluminescence gains in the linear region, followed by krypton, argon and neon. Ex\-tra\-po\-la\-ting the electroluminescence yield, we determined the electroluminescence threshold to be $1.51\pm0.04\textrm{ Td}$, $4.1\pm0.1\textrm{ Td}$, $2.6\pm0.1\textrm{ Td}$ and $2.9\pm0.1\textrm{ Td}$ for neon, argon, krypton and xenon, respectively, in good agreement with \cite{1d,GSPC-thvdias}. Neon, although it has the highest minimum excitation energy, has the lowest electroluminescence threshold. This is due to neon having the smallest elastic cross section of all gases studied. Thus, the electrons achieve higher energies at the same field.

\begin{figure}
\centering
\includegraphics[width=0.7\textwidth]{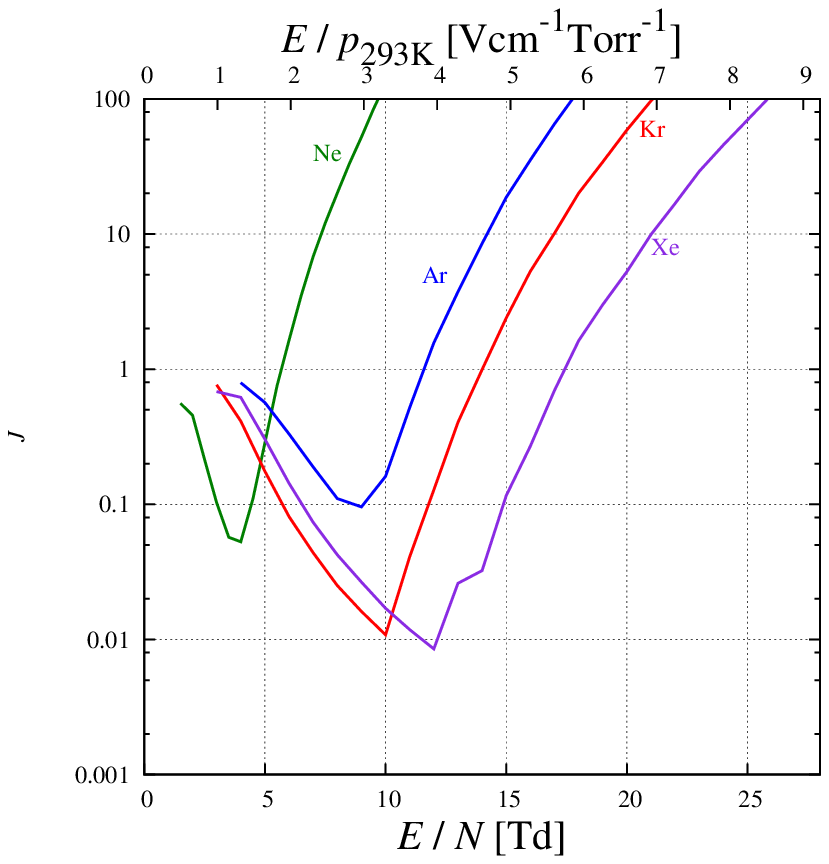}
\caption{Relative variance in the number of emitted photons as a function of $E/N$ for 2 cm of drift, 293 K and 1 atm.}
\label{fig:j}
\end{figure}

The relative variance in the number of emitted photons $J$, is defined as

\begin{equation}
\label{eq:j}
J=\frac{\sigma^2_{\mathrm{VUV}}}{\bar{N}_\mathrm{VUV}}
\end{equation}
where $\sigma^2_{\mathrm{VUV}}$ is the variance of the number of photons emitted by one primary electron, $N_\mathrm{VUV}$. This parameter, shown in Fig. \ref{fig:j}, is useful for determine the energy resolution of a detector \cite{en_res}. Xenon is the gas that exhibits the lowest statistical fluctuations, followed by krypton, argon and neon. For electric fields above the electroluminescence threshold, as the field increases, $J$ decreases until the onset of secondary electron production. At this point, $J$ begins to increase because the higher fluctuations in the charge gain start to dominate. To the best of our knowledge, up to now, there are not consistent values for $J$ published in literature.

\section{Conclusions}
\label{conclusions}

We used the new C++  version of the \textit{microscopic technique} of Garfield to access information on excited atoms produced during the electron drift through the pure noble gases neon, argon, krypton and xenon. The simulation relies on procedures and cross-sections available in Magboltz 8.9.3. We assumed that every excited atom leads to the emission of one VUV photon. We were able to calculate for a uniform electric field: the excitation and electroluminescence efficiencies, the electroluminescence yield and the statistical fluctuations as functions of the reduced electric field. 

The results obtained for the excitation and electroluminescence efficiencies and for the excitation thresholds are in good agreement with earlier Monte Carlo work. We also compared our results for the reduced electroluminescence yield with measurements for argon and xenon, and good agreement was found.

We simulated the statistical fluctuations associated with electroluminescence. As the reduced electric field increases, the statistical fluctuations decrease until the secondary charge fluctuations dominate. We confirm that the statistical fluctuations associated to proportional electroluminescence are lower than those in both charge avalanche multiplication and primary electron cloud formation. 

\section{Acknowledgments}
This work was partially supported by project CERN/FP/109283/2009 under the FCT (Lisbon) program. C. A. B. Oliveira was supported by FCT under Doctoral Grant SFRH/BD/36562/2007.


%% References


\bibliographystyle{elsarticle-num}
%%\bibliography{<your-bib-database>}

\begin{thebibliography}{00}


\bibitem{darkmatter1}
E. Aprile et al., Phys. Rev. Lett. \textbf{105} (2010) 131302.
\bibitem{darkmatter2}
http://lux.brown.edu [2011, March 15].
\bibitem{darkmatter3}
P. Benetti et al., Astropart. Phys. \textbf{28} (2008) 495.
\bibitem{next}
F. Gra\~{n}ena et al., 	arXiv:0907.4054v1 [hep-ex]. 
\bibitem{exo}
David Sinclair, ``EXO Gas: Ba tagging and tracking'', Xenon Detector Workshop, November 16-18, 2009, Berkeley, CA, USA.
\bibitem{3dxe}
F. P. Santos et al., J. Phys. D. Appl. Phys. \textbf{27} (1994) 42.
\bibitem{xe-ne}
F. P. Santos et al., IEEE T. Nucl. Sci. \textbf{45} (1998) 176.
\bibitem{1d}
T.H.V.T. Dias et al., J. Phys. D. Appl. Phys. \textbf{19} (1986) 527.
\bibitem{expxe}
C.M.B. Monteiro et al., J. Instrum. \textbf{2} (2007) P05001.
\bibitem{expar}
C.M.B. Monteiro et al., Phys. Lett. B \textbf{668} (2008) 167.
\bibitem{GSPC-thvdias}
T.H.V.T. Dias et al., J. Appl. Phys. \textbf{85} (1999) 6303. 
\bibitem{jl1}
I. I. Sobelman, ``Atomic Spectra and Radiative Transitions'', 2nd Edition, Springer-Verlag, 1992.
\bibitem{jl2}
G. Racah, Phys. Rev. \textbf{61} (1942) 537.
\bibitem{persistant1}
NIST, Basic Atomic Spectroscopic Data,\\
 http://www.nist.gov/physlab/data/handbook/index2.cfm [2011, March 15].
\bibitem{levels_ar}
Y. Ralchenko et al., 
NIST Atomic Spectra Database (ver. 4.0.0), http://www.nist.gov/physlab/data/asd.cfm  [2010, September 9].
\bibitem{ion}
http://physics.nist.gov/PhysRefData/IonEnergy/tblNew.html [2011, March 15].
\bibitem{persistant2}
NIST, Persistent Lines of Neutral Argon (Ar I), \\
http://physics.nist.gov/PhysRefData/Handbook/Tables/argontable3\_a.htm [2011, March 15].
\bibitem{suzuki}
M. Suzuki and S. Kubota, Nucl. Instrum. Methods \textbf{164} (1979) 197.
\bibitem{tanaka1}
Y. Tanaka, J. Opt. Soc. Am. \textbf{45} (1955) 710.
\bibitem{tanaka2}
Y. Tanaka et al., J. Opt. Soc. Am. \textbf{48} (1958) 304.
\bibitem{p2}
P. K. Leichner et al., Phys. Rev. A \textbf{13} (1976) 1787.
\bibitem{associative1}
M. S. B. Munson et al., J. Phys. Chem. \textbf{67} (1963) 1542.
\bibitem{associative2}
R. E. Huffman and D. H. Katayama, J. Chem. Phys. \textbf{45} (1966) 138.
\bibitem{mulliken}
R. S. Mulliken, J. Chem. Phys. \textbf{52} (1970) 5170.
\bibitem{koehler}
H. A. Koehler et al., Phys. Rev. A \textbf{9} (1974) 768.
\bibitem{fc1}
G. M. Barrow, ``Introduction to Molecular Spectroscopy'', McGraw-Hill International Editions, 1962.
\bibitem{fc2}
C. N. Banwell, ``Fundamentals of Molecular Spectroscopy'', 3rd edition, McGraw-Hill International Editions, 1972.
\bibitem{garf++}
http://cern.ch/garfieldpp [2011, March 15].
\bibitem{mag}
http://cern.ch/magboltz [2011, March 15].
\bibitem{garf_ex}
http://cern.ch/garfield/examples/ [2011, March 15].
\bibitem{nebem}
S. Mukhopadhyay and N. Majumdar, Eng. Anal. Bound. El. \textbf{33} (2009) 105.
\bibitem{mag_paper}
S.F. Biagi, Nucl. Instrum. Meth. A \textbf{421} (1999) 234.
\bibitem{Xe-meta}
M. Walhout et al., Opt. Lett. \textbf{20} (1995) 1192.
\bibitem{Ar-rad}
G. M. Lawrence, Phys. Rev. \textbf{175} (1968) 40.
\bibitem{Xe-rad}
D. K. Anderson, Phys. Rev. \textbf{137} (1965) A21.
\bibitem{morozov}
A. Morozov et al., J. Appl. Phys. \textbf{103} (2008) 103301.
\bibitem{saito}
K. Saito et al., IEEE T. Nucl. Sci. \textbf{49} (2002) 1674.
\bibitem{en_res}
J. M. F. dos Santos et al., X-Ray Spectrom. \textbf{30} (2001) 373.
\end{thebibliography}



\end{document}